\documentstyle[12pt]{article}
\newcommand {\beq}{\begin{equation}}
\newcommand {\eeq}{\end{equation}}
\newcommand {\beqa}{\begin{eqnarray}}
\newcommand {\eeqa}{\end{eqnarray}}
\newcommand {\beqan}{\begin{eqnarray*}}
\newcommand {\eeqan}{\end{eqnarray*}}

\newcommand {\Romannumeral}[1]{\uppercase\expandafter{\romannumeral#1}}

\newcommand {\ee}{\mbox{e}}

\newcommand {\dd}{\mbox{d}}

\newcommand {\del}{\partial}

\begin{document}
\setlength{\oddsidemargin}{0cm}

\begin{titlepage}
\renewcommand{\thefootnote}{\fnsymbol{footnote}}
    \begin{normalsize}
     \begin{flushright}
                 KEK-TH-401\\
                 UT-683\\
                 July 1994
     \end{flushright}
    \end{normalsize}
    \begin{Large}
       \vspace{1cm}
       \begin{center}
         {\Large Fractal Structure
in Two-Dimensional Quantum Regge Calculus} \\
       \end{center}
    \end{Large}

  \vspace{10mm}

\begin{center}
           Jun N{\sc ishimura}\footnote
           {E-mail address : nisimura@theory.kek.jp,
{}~JSPS Research Fellow.} {\sc and}
           Masaki O{\sc shikawa}\footnote
           {E-mail address : oshikawa@mmm.t.u-tokyo.ac.jp}\\
      \vspace{1cm}
        $\ast$ {\it National Laboratory for High Energy Physics (KEK),}\\
               {\it Tsukuba, Ibaraki 305, Japan}\\
        $\dagger$ {\it Department of Applied Physics, University of Tokyo,} \\
              {\it Bunkyo-ku, Tokyo 113, Japan}\\
\vspace{15mm}

\end{center}
\hspace{5cm}

\begin{abstract}
\noindent
We study the fractal structure of the surface in two-dimensional
quantum Regge calculus
by performing Monte Carlo simulation with up to 200,000 triangles.
The result can be compared with the universal scaling function
obtained analytically in the continuum limit of dynamical triangulation,
which provides us with a definite criterion
whether Regge calculus serves as a proper
regularization of quantum gravity.
When the scale-invariant measure is taken as the measure of the link-length
integration, we observe the correct scaling behavior
in the data for the type of loop attached to a baby universe.
The data seem to converge to the universal scaling function as the
number of triangles is increased. The data for the type of loop
attached to the mother universe, on the other hand, shows
no scaling behavior up to the present size.
\end{abstract}

\end{titlepage}

\vfil\eject

\setcounter{footnote}{0}

Construction of a consistent theory of quantum gravity has been
a long-standing problem in theoretical physics.
Although we have not reached any satisfactory conclusion yet
for the physically interesting four space-time dimensions,
there has been considerable progress
in two-dimensional quantum gravity \cite{2DQG}, which
is important not only as a toy model of four-dimensional quantum gravity,
but also as a prototype of theories of strings or random surfaces.
The progress is based on a continuum formulation --- Liouville theory
\cite{Liouville} --- and
a kind of lattice formulation --- dynamical triangulation
\cite{DT}.
Both have been exactly solved and their equivalence has been shown up to
the level of correlation functions \cite{equiv}.

In the case of ordinary quantum field theories,
continuum formulations and lattice formulations
have played complementary roles in understanding the universality
in field theory.
The details of lattice formulations at the lattice level are irrelevant to
the long-range behavior of the theory, which can be also described
in terms of a continuum formulation.
A natural explanation of the universality
has been given by the concepts of the renormalization group.
The equivalence of Liouville theory and
dynamical triangulation in two-dimensional quantum gravity
suggests that the universality exists also in quantum gravity.
In fact, some kinds of universality are known in dynamical triangulation;
the continuum limit is not affected by
using squares instead of triangles as the building blocks or
by prohibiting tadpoles or self-energies in the dual diagram, etc..
Although we still lack understanding of the universality in quantum gravity
in terms of the renormalization group,
invention of such a framework
(for some attempts, see \cite{BSTQG})
might be especially useful in studying quantum gravity
in three or four dimensions, where there is no analytic solution.

To this end, it seems quite important for us to investigate
the universality phenomena further in two-dimensional quantum gravity.
For example, there is a kind of
lattice formulation of quantum gravity based on Regge calculus~\cite{Regge}.
In this formulation, the lattice structure is fixed and the
integration over the metric is replaced by the integration over
the link lengths.
An interesting question to ask here is
whether two-dimensional Regge calculus falls into the same universality
class as dynamical triangulation and Liouville theory.
This is quite non-trivial,
since it is questionable whether
the general covariance arises in the continuum limit of such a system
that can be viewed as a
statistical model on a regular lattice.
The answer is also desired from a practical point of view,
since Regge calculus
might be useful for numerical simulations of
quantum gravity in higher dimensions.

Since analytic treatment of Regge calculus seems difficult beyond
a perturbative expansion around flat space-time \cite{wcexpansion},
we investigate the above issue by numerical simulation.
There have been several works in this direction.
A few years ago, Gross and Hamber~\cite{GrossHamber}
reported that Regge calculus reproduced
the string susceptibility known in dynamical
triangulation and in the continuum theory.
Recently, Bock and Vink~\cite{BockVink}
have performed a more careful analysis
and have claimed that Regge calculus
fails to reproduce the desired string susceptibility.
It has been also reported~\cite{GrossHamber,HolmJanke} that
the critical exponents of the Ising model on the dynamical Regge lattice
agree quite well with the ones of the Ising model on the static lattice
(Onsager's values)
and not with the ones of the Ising model on the
dynamically triangulated lattice.
We think, however, that all these works are subject to
some subtleties in Regge calculus
concerning either the definition of string susceptibility
or the introduction of matter fields, as we explain later.

In order to compare Regge calculus with the other approaches
unambiguously,
it is desirable to have a universal quantity
which can be calculated directly from the geometry of the surface.
Indeed, such a quantity exists; it is the so-called
loop-length distribution, which has been studied by
Kawai, Kawamoto, Mogami and Watabiki~\cite{KKMW}.
They constructed a transfer-matrix formalism in dynamical triangulation.
Using the formalism, they succeeded in obtaining the loop-length
distribution in the continuum limit,
which characterizes the fractal structure of the surface.
The loop-length distribution can be defined unambiguously also in
Regge calculus and it can be measured through numerical simulation.
We, therefore, examine the loop-length distribution in
a numerical simulation of Regge calculus and compare the
result with that obtained in the continuum limit of dynamical triangulation.

\vspace{1cm}

Let us first explain the system we consider in this letter.
In Regge calculus the dynamical variables are the link lengths
on a fixed triangulation.
Since it is essential for our purpose to have a spherical topology,
we construct the fixed triangulation by dividing
each of the twenty surfaces of an icosahedron into a triangular lattice.
A tetrahedron or octahedron could be used as well,
but we have chosen an icosahedron
so that the artifact of the non-uniformity of coordination number may be
the minimum.
The integration over the link lengths is performed taking either
the uniform measure
$\int \prod_i \dd l_i$ or the scale-invariant measure
$\int \prod_i \dd l_i/l_i$. The triangle inequality is imposed
on every triangle.
The total area is fixed to be equal to the number of triangles, so that the
average area of a triangle in each configuration becomes unity.
This can be realized in the simulation
as follows.
We first consider a system with the measure $\int \prod_i \dd l_i ~ l_i
^{~p-1}$
and the action $S=\lambda A$, where $A$ is the total area of the surface
and $\lambda$ is a constant parameter.
$p=1$ corresponds to the uniform measure and $p=0$ corresponds to the
scale-invariant measure.
The partition function is given by
\beq
Z(\lambda)=\int \prod_i \dd l_i~ l_i^{~p-1}~\ee ^{-\lambda A}
{}~\Theta(\mbox{\it triangle inequalities}),
\label{eq:zlambda}
\eeq
where $\Theta(\mbox{\it triangle inequalities})$
is the step function which gives
one if all the triangle inequalities are satisfied and zero otherwise.
Changing the variables of integration as $l_i \rightarrow l_i/\sqrt{\lambda}$,
one can extract the explicit $\lambda$-dependence of $Z(\lambda)$ as
\beq
Z(\lambda) = \frac{1}{\left(\sqrt{\lambda}\right)^{pN_{\mbox{\tiny link}}}}
\int \prod_i \dd l_i ~ l_i ^{~p-1}~ \ee ^{-A}~
\Theta(\mbox{\it triangle inequalities}),
\eeq
where $N_{\mbox{\scriptsize link}}$ is the number of links.
Inverse Laplace transform of eq. (\ref{eq:zlambda})
gives the distribution of the total area as
\beq
\varphi(A) \propto \ee ^{-\lambda A} A ^{\frac{p N_{\mbox{\tiny link}}}{2}-1}.
\label{eq:area}
\eeq
For the uniform measure ($p=1$), we control the total area by
introducing a positive $\lambda$, while
for the scale-invariant measure ($p=0$), we set $\lambda=0$ and
rescale the configuration whenever necessary during the simulation
so that the total area may be kept within a range of moderate value.
The configurations thus generated for either measure are
each rescaled before measurements
to have the fixed total area equal to the number of triangles.
We have performed Monte Carlo simulations using the heat-bath algorithm
with 50,000 triangles for the uniform measure and
with 12,500, 50,000 and 200,000 triangles for the scale-invariant measure.
The updating process in the program is vectorized since we can update
one third of the links independently at the same time.

Before proceeding, we would like to explain the subtleties in Regge calculus.
The first point is the definition of string susceptibility.
In dynamical triangulation, the string susceptibility
$\gamma_{\mbox{\scriptsize str}}$
can be defined through
\beq
 Z(N)=\sum_{T\in {\cal T}_N} \ee^{-S(T)} \sim N^{\gamma_{\mbox{\tiny str}}-3}
\ee^{\kappa N}~~~~~~~~(N \rightarrow \infty),
\label{eq:KPZ}
\eeq
where ${\cal T}_N$ denotes the set of triangulations with $N$
triangles~\cite{Ambjorn}.
In Regge calculus, the partition function for a fixed triangulation with
$N$ triangles can be written as
\beq
Z(A,N)=\int \dd \mu(\{l_i\}) ~\ee ^{-S(\{l_i\})}
{}~\delta(A(\{l_i\})-A)~\Theta(\mbox{\it triangle inequalities}),
\eeq
where $\dd\mu(\{l_i\})$ denotes the measure for the link-length integration,
and $A(\{l_i\})$ denotes the total area of the surface, which is fixed to
a given value $A$.
In refs. \cite{GrossHamber,BockVink}, they defined the string susceptibility
in Regge calculus through
\beq
Z(A,N)\sim A^{\gamma_{\mbox{\tiny str}}-3}
\ee^{\kappa A}~~~~~~~~~~(A \rightarrow \infty),
\eeq
for a fixed $N$, which is taken to be sufficiently large.
This definition of string susceptibility, however,
might be too naive.
When we consider a scaling relation in field theory, we have to keep
the cutoff of the theory constant.
In Regge calculus, we have no definite quantity that corresponds to the cutoff
in ordinary field theory,
and therefore the scaling argument is rather subtle.
One natural thing to do is to consider the average area
of a triangle in each configuration as the cutoff in ordinary field theory.
In order to keep the cutoff constant, say at unity,
we should fix the total area $A$ to be equal to the number of triangles $N$.
The string susceptibility, then, can be defined through
\beq
 Z(A=N,N)\sim N^{\gamma_{\mbox{\tiny str}}-3}
\ee^{\kappa N}~~~~~~~~~~(N \rightarrow \infty).
\label{eq:correct}
\eeq
Unfortunately, the string susceptibility thus defined
seems to be difficult
to extract from numerical simulation, since we have to probe the difference
in free energy for different numbers of triangles.
The second point is the introduction of matter fields.
If we assign a single spin to each triangle,
each spin should be regarded as a representative~(``block spin'') of
the dynamical degrees of freedom within the triangle.
Therefore, we may have to, for example, make the Ising coupling
constant dependent on the
size of the triangle, which takes a different value from point to point
in Regge calculus.
Thus the negative results obtained in ref.~\cite{GrossHamber,HolmJanke}
might be due to the problem of the action for the matter fields.
In contrast to the above points, the loop-length distribution
can be defined unambiguously in Regge calculus and
we hope this provides us with a definite criterion
whether Regge calculus serves as a proper regularization of quantum gravity.

Let us explain the loop-length distribution, which plays a central
role in our study.
The set of points which are at a distance $D$ from a given point is
composed of a number of disconnected closed loops.
The distribution $\rho(L,D)$ of the loop length $L$ at the distance $D$
has been obtained in the continuum limit of dynamical triangulation as
\cite{KKMW}
\beqa
\rho(L,D)&=&\frac{1}{D^2} f(x) \\
f(x)&=&\frac{3}{7\sqrt{\pi}} \left(
x^{-\frac{5}{2}}+ \frac{1}{2} x^{-\frac{3}{2}}+\frac{14}{3} x^{\frac{1}{2}}
\right) \ee ^{-x},
\eeqa
where $x=L/D^2$.
The fact that such a quantity does possess a sensible continuum limit
is not only quite non-trivial itself but also of great significance
since it provides us with a geometrical picture of the continuum limit of
quantum gravity.
It implies that in the continuum limit of quantum gravity,
the space-time becomes fractal
in the sense that sections of the surface at different distances from
a given point look exactly the same after a proper rescaling of loop lengths.
Let us here consider the distribution of
loops with length $L$ attached to a universe with area $A'$.
This can be written in terms of the functions given in ref. \cite{KKMW} as,
\beq
\rho(L,A';D;A)=
\lim_{L_0 \rightarrow 0}
\frac{\frac{1}{2\pi i}\int \dd \tau ' N(L_0,L;D;\tau ')\ee ^{\tau ' (A-A')}
\frac{1}{2\pi i}\int \dd \tau '' \frac{1}{L} F(L,\tau '') \ee ^{\tau'' A'}}
{\frac{1}{2\pi i}\int \dd \tau '' \frac{1}{L_0} F(L_0,\tau '') \ee ^{\tau''
A}},
\eeq
where $A$ is the total area of the surface.
When we take the thermodynamic limit $A\rightarrow\infty$, we can
fix either $A'$ or $(A-A')$ to a finite value $B$.
The former corresponds to the type of loop attached to a baby universe,
while the latter corresponds to the type of loop attached to the mother
universe.
We refer to the two types of loops simply
as ``baby'' loops and ``mother'' loops,
respectively (fig.~1).
Note that baby loops are not necessarily smaller than mother loops.
The distribution for baby loops can be calculated as follows,
\beqa
\rho_{\mbox{\scriptsize b}}(L,B;D)
&=&\frac{1}{2\pi i} \int \dd \tau ' \ee ^{\tau ' B}
   \lim_{L_0\rightarrow 0,\tau \rightarrow 0}
    \frac{\frac{\del^2}{\del \tau ^2} N(L_0, L;D;\tau)
     \frac{1}{L} F(L,\tau+\tau')}
    {\frac{\del^2}{\del \tau ^2} \frac{1}{L_0} F(L_0,\tau)} \\
&=&\frac{1}{D^6} \frac{6}{7\pi} \left(1+\frac{1}{2} x \right) \ee^{-x}
   x^{\frac{1}{2}} y^{-5} \ee ^{-\left(\frac{x}{y}\right)^2},
\eeqa
where $x=L/D^2$ and $y=\sqrt{B}/D^2$.
Integrating over $B$, one gets,
\beqa
\rho_{\mbox{\scriptsize b}}(L,D)&=&
\int_0^{\infty}  \rho_{\mbox{\scriptsize b}}(L,B;D) \dd B  \\
&=&  \frac{1}{D^2} f_{\mbox{\scriptsize b}}(x),
\eeqa
where
\beq
f_{\mbox{\scriptsize b}}(x)= \frac{3}{7\sqrt{\pi}}
\left( x^{-\frac{5}{2}} + \frac{1}{2} x^{-\frac{3}{2}} \right) \ee^{-x}.
\label{eq:baby}
\eeq
The distribution for mother loops can be calculated similarly.
\beqa
\rho_{\mbox{\scriptsize m}}(L,B;D)
&=&\frac{1}{2\pi i} \int \dd \tau ' \ee ^{\tau ' B}
   \lim_{L_0\rightarrow 0,\tau\rightarrow 0}
    \frac{\frac{\del^2}{\del \tau ^2} N(L_0, L;D;\tau+\tau')
     \frac{1}{L} F(L,\tau)}
    {\frac{\del^2}{\del \tau ^2} \frac{1}{L_0} F(L_0,\tau)} \\
&=& \frac{1}{\sqrt{\pi L}} \frac{\del}{\del D}
\left[ \frac{1}{2\pi i} \int _{-i\infty} ^{i \infty}
\dd \tau \ee ^{\tau B} \exp \left\{ - \frac{\sqrt{\tau}L}{2} \left(3\tanh^{-2}
\frac{\sqrt{6\sqrt{\tau}}D}{2}-2 \right) \right\}   \right].
\eeqa
Integrating over $B$, one gets,
\beqa
\rho_{\mbox{\scriptsize m}}(L,D)&=&
\int_0^{\infty}  \rho_{\mbox{\scriptsize m}}(L,B;D)\dd B \\
&=&  \frac{1}{\sqrt{\pi L}} \frac{\del}{\del D}
\lim_{\tau\rightarrow 0} \exp \left\{
-\frac{\sqrt{\tau}L}{2} \left(3 \tanh^{-2}
\frac{\sqrt{6\sqrt{\tau}}D}{2}-2 \right) \right\} \\
&=&\frac{1}{D^2} f_{\mbox{\scriptsize m}}(x),
\eeqa
where
\beq
 f_{\mbox{\scriptsize m}}(x)
 = \frac{2}{\sqrt{\pi}}  x^{\frac{1}{2}} \ee^{-x}.
\label{eq:mother}
\eeq
The total distribution reproduces the result of ref. \cite{KKMW}
\footnote{The fact that the loop-length distribution is composed of two
such contributions can also be seen from eq.(24) of ref. \cite{KKMW}.
One can get the singularity $\tau^{3/2}$ in the numerator
either from the proper-time evolution
kernel $N(L_0,L;D)$ or from the disk amplitude $F(L)$.
The former contribution corresponds to
baby loops, while the latter to mother loops \cite{Kawai}.},
{\it i.e.} $\rho(L,D)= \rho_{\mbox{\scriptsize b}} (L,D)
+\rho_{\mbox{\scriptsize m}}(L,D)$.
Note that the integration of the mother-loop distribution
over $L$ gives
\beq
\int_0 ^{\infty}   \rho_{\mbox{\scriptsize m}}(L,D) \dd L
=\frac{2}{\sqrt{\pi}} \int_0^{\infty} x^{\frac{1}{2}} \ee ^{-x} \dd x =1,
\eeq
which means that one finds exactly one mother
loop at a fixed $D$ in each configuration.

In Regge calculus we define the loop-length distribution
in the following way.
Representing each triangle by the center of its inscribed circle,
we define the length of a dual-link
by the geodesic distance of the
two representative points at the ends of the dual-link.
Then we define the distance between two given triangles by
the length of the shortest dual-link path
connecting the two triangles (fig.~2).
The boundary between triangles at a distance less than or equal to $D$
from a given triangle and those
at a distance greater than $D$ from the same triangle is
composed of disconnected closed loops, whose lengths are defined
by summing up the lengths of the links forming each loop.
This gives the definition of the loop-length distribution in Regge calculus.
Since we are dealing with a finite number of triangles, we have a maximum $D$
at which all the triangles are included.
We measure the loop-length distribution at $D$'s which are
less than half of the maximum $D$.
Also the finiteness of the system makes
the classification of loops into baby loops and mother loops
somewhat ambiguous.
We identify the mother loop with the loop attached
to the largest universe at a fixed $D$.
The measurement has been made every 200 sweeps, and
at each measurement we choose ten triangles randomly as the starting point
of the distance $D$ in order to increase the statistics.
The number of sweeps required for the thermalization of the data
is large for the scale-invariant measure; {\it e.g.} 500,000 sweeps
in the case of 12,500 triangles.

\vspace{1cm}

Let us show the results of our simulation.
The data for the uniform measure ($\int \prod_{i} d l_i$) with
50,000 triangles is shown in Fig. 3.
The loop-length distribution
is plotted against the scaling variable $x=L/D^2$
for $D=20,40,60,80,100,120$, where
$L$ is the length of the loop and $D$ is the distance from
a point on the surface.
The result does not show any scaling behavior in terms of
$x$ for different values
of $D$.
The distribution for each $D$ is split into two parts:
the left one which
corresponds to baby loops and the right one which corresponds to mother loops.
One should note that it is a log-log plot, and so we see that
the baby loops are extremely suppressed.
Indeed, for most cases we find only one loop, which is the mother loop, for
each $D$ in a configuration.
(A baby loop appears on average only once
in five configurations even for $D=120$.)
This means that the surface is rather similar to
a smooth sphere, where always only one loop (which is the mother loop)
appears.
Hence the surface is quite different from
being fractal as in dynamical triangulation,
where many loops appear for each $D$.

The data for the scale-invariant measure
($\int \prod_{i} d l_i /l_i$)
with 12,500 triangles is
shown in Fig. 4.
The loop-length distribution
is plotted against the scaling variable $x=L/D^2$
for $D=10,15,20$.
Something of a scaling behavior is seen in the intermediate region of $x$.
To clarify the situation, we separate the two contributions,
the one from
baby loops and the one from mother loops.
The distribution for baby loops is shown in Fig. 5.
A clear scaling behavior can be seen with $x=L/D^2$ as a scaling parameter.
Furthermore, we compare our result with the universal function (\ref{eq:baby})
obtained in the continuum limit of dynamical triangulation.
Since there is an ambiguity by a constant factor between the
scaling parameter $x$ in our system and that in eq. (\ref{eq:baby}),
we fit our result with $\alpha f_{\mbox{\scriptsize b}}(\alpha x)$,
where $\alpha$ is the
fitting parameter.
The best fit ($\alpha=3.2$) is shown by the dotted line in Fig. 5.
Our data is in good agreement with the universal function.
Let us see how the slight discrepancy seen
in the small-$x$ region behaves as we increase the number of triangles
$N$.
Fig. 6 shows the results
for $D=20$ with 12,500, 50,000 and 200,000 triangles.
We find that the data curve in the small-$x$ region
approaches to the curve of the universal
function (\ref{eq:baby}).
Thus we expect that the distribution of baby loops will converge
to the universal function in the
$N\rightarrow \infty$ limit.
The distribution for mother loops, on the other hand,
is shown in Fig. 7 and Fig. 8, which correspond to the cases with
12,500 triangles and 200,000 triangles respectively.
We cannot see any scaling behavior here.
The dotted line represents the rescaled universal function
$\alpha f_{\mbox{\scriptsize m}}(\alpha x)$
with the same $\alpha=3.2$ that gives the best fit in the case of the baby-loop
distribution.
Although we might expect that the data will approach the universal function
for larger $D$ with sufficiently many triangles,
the finite-size effect in the present data is too severe for us
to draw any conclusion.

\vspace{1cm}

To summarize,
we have performed a Monte Carlo simulation of two-dimensional Regge calculus
up to 200,000 triangles
and measured the loop-length distribution, which is the distribution
function of the length $L$ of the loops whose geodesic distance
from a point is a constant $D$.
The results are compared with that obtained in the continuum limit of dynamical
triangulation, which has the scaling behavior in terms of $x = L/D^2$.

For the uniform measure ($\int \prod_{i} dl_i$), we find no scaling
in terms of $x$ at the present size.
Moreover, we find for most cases only one loop for each $D$.
This means that the surface is smooth, in contrast to the fractal structure
in dynamical triangulation.

For the scale-invariant measure ($\int \prod_{i} \dd l_i / l_i$),
we find that the baby-loop distribution
shows a clear scaling behavior in terms of $x$, which is in good agreement
with the universal function obtained in the continuum limit of dynamical
triangulation.
It is rather surprising that
Regge calculus and dynamical triangulation, which seem to
be quite different systems, show the same universal behavior at least
for the baby-loop distribution.
It is not clear at present whether the scale-invariant measure is
essential in obtaining this universal behavior.
One might expect that if there is a universality at all, the choice of
the measure of the link-length integration would be irrelevant to the
universal behavior.
We should note, however, that the scale-invariant measure is very special,
as is seen in eq. (\ref{eq:area}).
It implies that the fluctuation of the area of each triangle becomes
overwhelmingly large when the scale-invariant measure is adopted.
Since the lattice structure is regular in Regge calculus,
such a large fluctuation might be necessary to obtain a fractal structure
as in dynamical triangulation.

As for the mother-loop distribution, the situation is not clear even for the
scale-invariant measure.
We feel it is possible that the mother-loop distribution,
as well as the baby-loop distribution,
converges to the universal function
in the $N\rightarrow \infty$ limit.
Even if this turns out to be the case, we may at least say that
Regge calculus requires many more triangles in order to obtain the universal
behavior than dynamical triangulation, which is known to show
a fairly good universal fractal structure with no more than 20,000
triangles \cite{TsudaYukawa}.

\vspace{1cm}

We would like to thank H. Kawai for stimulating discussions and
for continuous encouragement.
It is our pleasure to acknowledge M. Ninomiya, M. Fukuma
and M. Oogami for useful comments.
We are also grateful to T. Yukawa and N. Tsuda for providing us
with their unpublished data of loop-length distribution in two-dimensional
dynamical triangulation and to N. McDougall for carefully reading the
manuscript.
The calculations were carried out on the supercomputer
HITAC S820/80 at National Laboratory for High Energy Physics(KEK).

\newpage

\centerline{\Large Figure captions}
\bigskip
\noindent Fig. 1
The loops appearing at the distance $D$ from a given point.
There are two types of loops : the ``mother'' loop which is attached
to the mother universe, and ``baby'' loops which are attached to
baby universes.

\bigskip
\noindent Fig. 2
An example of a dual-link path connecting the two shaded triangles.
The distance between the two triangles is defined by
the length of the shortest dual-link path.

\bigskip
\noindent Fig. 3
The loop-length distribution for the uniform measure ($\int \prod_i \dd l_i$)
with 50,000 triangles. The horizontal axis is the scaling variable
$x=L/D^2$.

\bigskip
\noindent Fig. 4
The loop-length distribution for the scale-invariant measure
($\int \prod_i \dd l_i/l_i $)
with 12,500 triangles. The horizontal axis is the scaling variable
$x=L/D^2$. The dotted curve is
the rescaled universal function
$\alpha f(\alpha x)$
with the same $\alpha=3.2$ that gives the best fit in the case of the baby-loop
distribution (Fig. 5).

\bigskip
\noindent Fig. 5
The baby-loop length distribution for the scale-invariant measure
($\int \prod_i \dd l_i/l_i$)
at $D=10,15,20$ with 12,500 triangles.
The horizontal axis is the scaling variable
$x=L/D^2$.
The dotted curve is
the rescaled universal function
$\alpha f_{\mbox{\scriptsize b}}(\alpha x)$
with $\alpha=3.2$ which gives the best fit.

\bigskip
\noindent Fig. 6
The baby-loop length distribution for the scale-invariant measure
($\int \prod_i \dd l_i/l_i$)
at $D=20$ with 12,500, 50,000, 200,000 triangles.
The horizontal axis is the scaling variable
$x=L/D^2$.

\bigskip
\noindent Fig. 7
The mother-loop length distribution for the scale-invariant measure
($\int \prod_i \dd l_i/l_i$)
at $D=10,15,20$
with 12,500 triangles. The horizontal axis is the scaling variable
$x=L/D^2$. The dotted curve is
the rescaled universal function
$\alpha f_{\mbox{\scriptsize m}}(\alpha x)$
with the same $\alpha=3.2$ that gives the best fit in the case of the
baby-loop distribution.

\bigskip
\noindent Fig. 8
The mother-loop length distribution for the scale-invariant measure
($\int \prod_i \dd l_i/l_i$)
at $D=20,40,60$
with 200,000 triangles. The horizontal axis is the scaling variable
$x=L/D^2$.

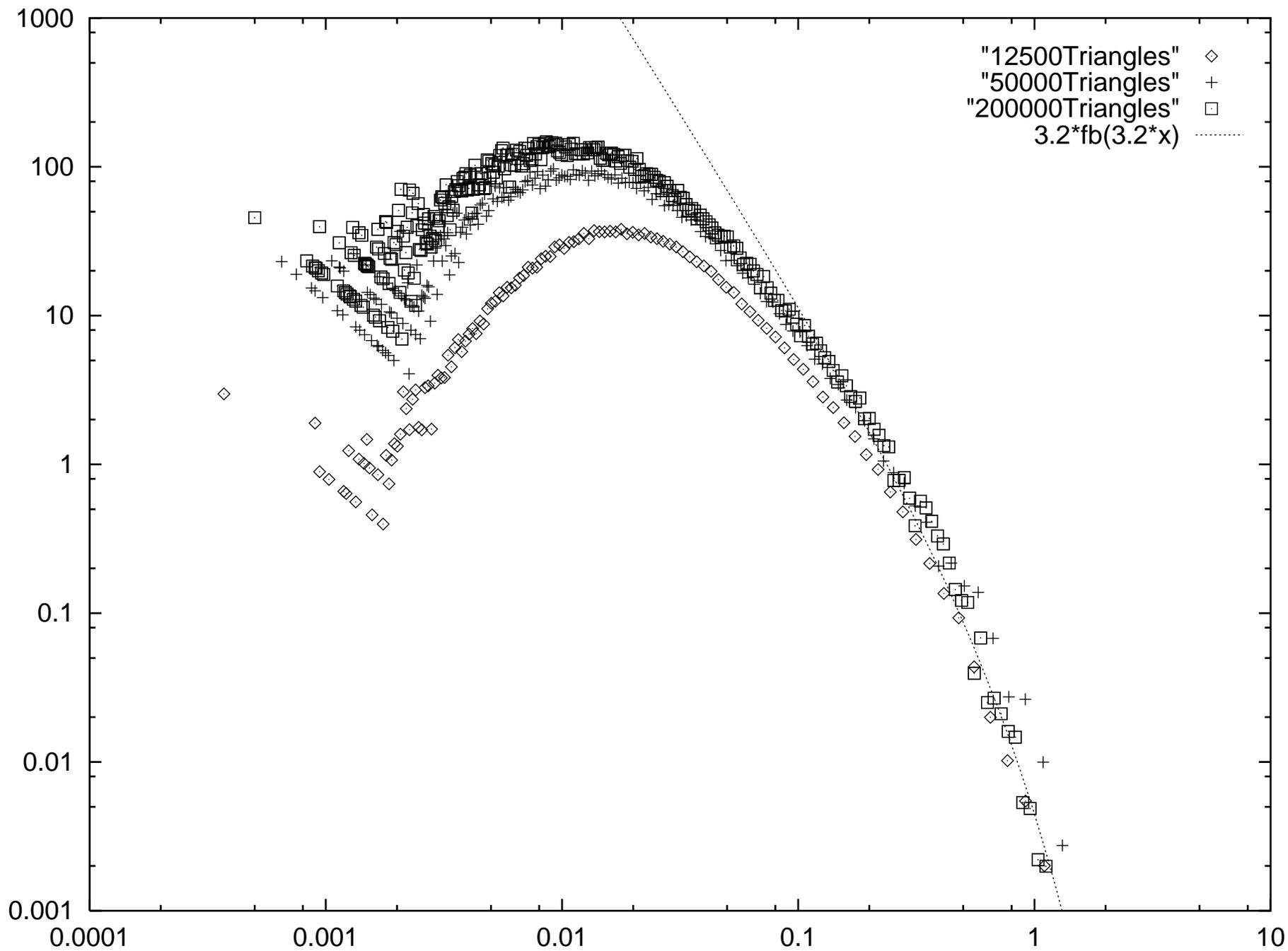
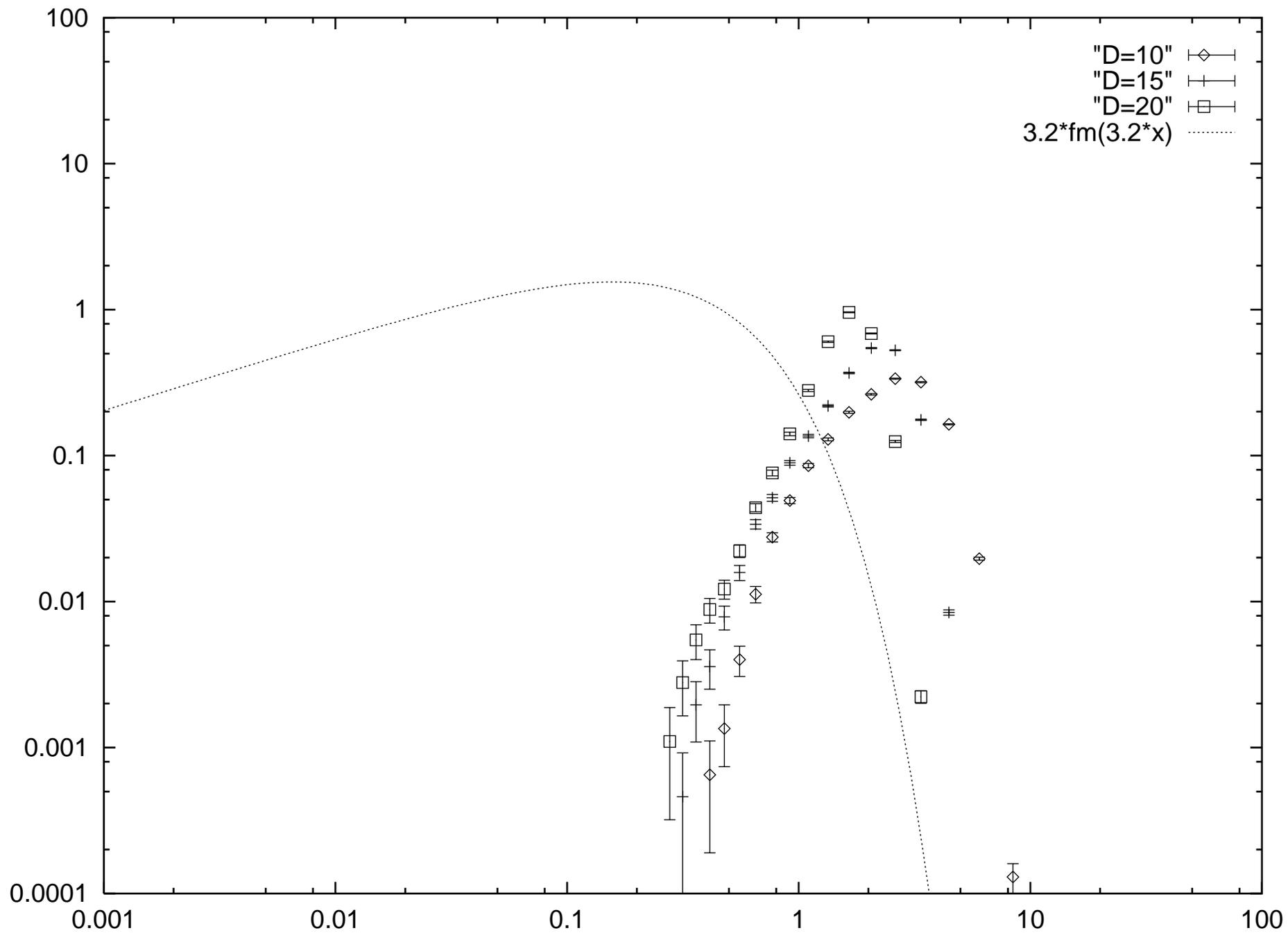
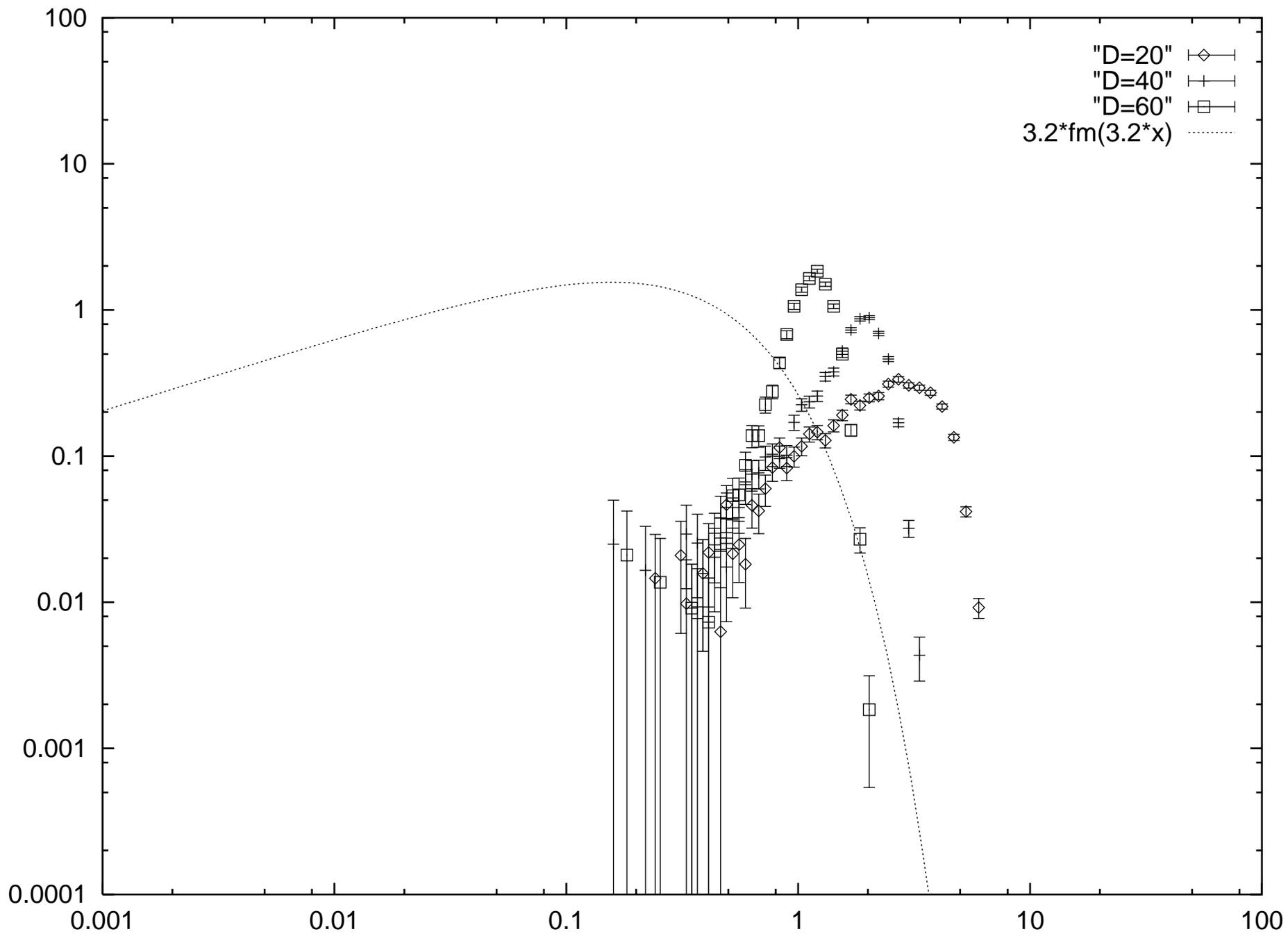
\end{document}